\documentclass[aps,prl,superscriptaddress,reprint,floats,citeautoscript,floatfix,nobibnotes,nofootinbib]{revtex4-1}

\usepackage[intlimits,tbtags]{amsmath}
\usepackage{times}
\usepackage{color}
\usepackage{amssymb}
\usepackage{graphicx}
\usepackage{bm}
\usepackage{array}
\usepackage{dcolumn}
\usepackage{bm}
\usepackage{setspace}

\newcolumntype{d}[1]{D{.}{.}{#1} }
\usepackage{xcolor}

\begin{document}

\title{Creating cyclo-N$_5$$^{+}$ cation and assembling N$_5$$^{+}$N$_5$$^{-}$ salt via electronegativity co-matching in tailored ionic compounds}

\author{Bi Zhang}\thanks{These authors contributed equally to this work.}
\affiliation{Laboratory of Quantum Functional Materials Design and Application, School of Physics and Electronic Engineering, Jiangsu Normal University, Xuzhou 221116, China}

\author{Yu Xin}\thanks{These authors contributed equally to this work.}
\affiliation{Key Laboratory of Material Simulation Methods and Software of Ministry of Education and State Key Laboratory of Superhard Materials, College of Physics, Jilin University, 130012, Changchun, China}

\author{Meiling Xu}\email{xml@calypso.cn}
\affiliation{Laboratory of Quantum Functional Materials Design and Application, School of Physics and Electronic Engineering, Jiangsu Normal University, Xuzhou 221116, China}
\affiliation{Key Laboratory of Material Simulation Methods and Software of Ministry of Education and State Key Laboratory of Superhard Materials, College of Physics, Jilin University, 130012, Changchun, China}

\author{Yiming Zhang}
\affiliation{Laboratory of Quantum Functional Materials Design and Application, School of Physics and Electronic Engineering, Jiangsu Normal University, Xuzhou 221116, China}

\author{Yinwei Li}\email{yinwei$_$li@jsnu.edu.cn}
\affiliation{Laboratory of Quantum Functional Materials Design and Application, School of Physics and Electronic Engineering, Jiangsu Normal University, Xuzhou 221116, China}

\author{Yanchao Wang}\email{wyc@calypso.cn}
\affiliation{Key Laboratory of Material Simulation Methods and Software of Ministry of Education and State Key Laboratory of Superhard Materials, College of Physics, Jilin University, 130012, Changchun, China}

\author{Changfeng Chen}\email{changfeng.chen@unlv.edu}
\affiliation{Department of Physics and Astronomy, University of Nevada, Las Vegas, NV, 89154, USA}

\begin{abstract}
The recent discovery of crystalline pentazolates marks a major advance in polynitrogen science and raises prospects of making the long-touted potent propellant N$_5$$^{+}$N$_5$$^{-}$ salt. However, despite the synthesis of cyclo-N$_5$$^{-}$ anion in pentazolates, counter cation cyclo-N$_5$$^{+}$ remains elusive due to the strong oxidizing power of pentazole ion; moreover, pure N$_5$$^{+}$N$_5$$^{-}$ salt is known to be unstable. Here, we devise a new strategy for making rare cyclo-N$_5$$^{+}$ cation and assembling the long-sought N$_5$$^{+}$N$_5$$^{-}$ salt in tailored ionic compounds, wherein the negative/positive host ions act as oxidizing/reducing agents to form cyclo-N$_5$$^{+}$/N$_5$$^{-}$ species. This strategy is implemented via an advanced computational crystal structure search, which identifies XN$_5$N$_5$F (X = Li, Na, K) compounds that stabilize at high pressures and remain viable at ambient pressure-temperature conditions based on \textit{ab initio} molecular dynamics simulations. This finding opens an avenue for creating and stabilizing N$_5$$^{+}$N$_5$$^{-}$ salt assembly in ionic compounds, where cyclo-N$_5$ species are oxidized/reduced via co-matching with host ions of high/low electronegativity. The present results demonstrate novel polynitrogen chemistry, and these findings offer new insights and prospects in the design and synthesis of diverse chemical species that exhibit unusual charge states, bonding structures, and superior functionality.  
  
\end{abstract}
\maketitle

\section{Introduction}

Nitrogen is highly abundant on Earth, constituting $\sim$78$\%$ of the atmosphere in diatomic gas form. At cryogenic temperatures, nitrogen undergoes a phase transition and solidifies into a crystalline structure comprised of N$_2$ molecules. Under broad temperature-pressure conditions, nitrogen adopts a wide spectrum of structural forms, including polyatomic molecules such as cyclic N$_3$ ~\cite{samartzis2005two}, and N$_4$ ~\cite{cacace2002experimental}, molecular crystals comprising N$_{x}$ (\textit{x} = 2 ~\cite{bini2000high,olijnyk1990high,turnbull2018unusually,laniel2023structure}, 6 ~\cite{greschner2016new}, 8 \cite{hirshberg2014calculations}, 10 \cite{liu2020novel}, and 12 \cite{lin2022ambient}, and N$_2$+N$_6$ \cite{mattson2004prediction}) units, such as \textit{$\alpha$}-N, \textit{$\varepsilon$}-N, \textit{$\iota$}–N$_2$, \textit{$\zeta$}-N$_2$, etc., amorphous structures \cite{eremets2001semiconducting,goncharov2000optical}, and polymeric solids exemplified by cg-N \cite{sun2013stable,laniel2019hexagonal,ji2020nitrogen,laniel2020high,wang2012cagelike,ma2009novel,tomasino2014pressure,eremets2004single}. Crystalline polynitrogen molecules and polymeric nitrogen are high-energy density materials with non-toxic decomposition products, making them highly appealing in many applications.

Meanwhile, polynitrogen ions, such as the linear azide N$_3$$^{-}$ anion \cite{curtius1890ueber}, the V-shaped N$_5$$^{+}$ cation \cite{christe1999n5+,wilson2003polynitrogen,haiges2004high,christe2007recent}, the pentazolate cyclo-N$_5$$^{-}$ anion \cite{vij2002experimental,bazanov2016detection}, and N$_8$$^{-}$ anion \cite{wu2014n8}, constitute an important class of structural forms in the field of polynitrogen chemistry. The N$_3$$^{-}$ and N$_5$$^{+}$ ions have been recognized and studied for many years, and the first discovery of these ions dates back to 1890 for N$_3$$^{-}$ and 1999 for N$_5$$^{+}$. The pentazolate anion, cyclo-N$_5$$^{-}$, has garnered more significant attention as a promising energetic polynitrogen species due to its extensive existence in a wide variety of compounds \cite{zhang2017synthesis,xu2017series,xu2017carbon,zhang2017symmetric,xu2018self,peng2015crystalline,shen2015novel,steele2016sodium,steele2017novel,williams2017novel,peng2015exotic,li2018simple,zhu2016stable,liu2020moderate,xia2019pressure,yuan2021high,xia2019predictions,zhou2020lithium,laniel2018high,bykov2021stabilization,sui2023high,steele2017high}. In 2017, cyclo-N$_5$$^{-}$ anion was first synthesized in solid (N$_5$)$_6$(H$_3$O)$_3$(NH$_4$)$_4$Cl salt at atmospheric conditions \cite{zhang2017synthesis}, and subsequently identified in various additional pentazolate hydrate complexes such as [Na(H$_2$O)(N$_5$)$_2$]$\cdot$2H$_2$O, [Mg(H$_2$O)$_6$(N$_5$)$_2$]$\cdot$4H$_2$O, and [M(H$_2$O)$_4$(N$_5$)$_2$]$\cdot$4H$_2$O, where M represents Mn, Fe, Co and Zn \cite{xu2017series,xu2017carbon,zhang2017symmetric}, as well as a three-dimensional metal-organic framework [Na$_8$(N$_5$)$_8$(H$_2$O)$_3$]$_{n}$ \cite{xu2018self}. Theoretical calculations have indicated that cyclo-N$_5$$^{-}$ anion can exist in pressure-stabilized simple metal pentazolates, such as MN$_5$ \cite{peng2015crystalline,shen2015novel,steele2016sodium,steele2017novel,williams2017novel,peng2015exotic,li2018simple,zhu2016stable}, M(N$_5$)$_2$ \cite{liu2020moderate,xia2019pressure,yuan2021high}, and M(N$_5$)$_3$ \cite{xia2019predictions}, where M refers to metal cations. Notably, alkali-metal pentazolate salts, including LiN$_5$ \cite{zhou2020lithium,laniel2018high}, NaN$_5$ \cite{bykov2021stabilization}, KN$_5$ \cite{sui2023high}, and CsN$_5$ \cite{steele2017high}, have been successfully realized in experiments. Of particular interest is the potential retrieval of LiN$_5$ at ambient conditions \cite{laniel2018high}.

Concerted efforts have been dedicated to exploring new polynitrogen species, both computationally and in the laboratory \cite{laniel2019synthesis,bykov2021high,zhai2022stabilized,bykov2022stabilization,salke2021tungsten,wang2022stabilization,laniel2023aromatic,bykov2018fe,bykov2020high,bykov2018high,laniel2021high,aslandukov2021high}. Conspicuously absent, however, is the counter cation of cyclo-N$_5$$^{-}$, cyclo-N$_5$$^{+}$, which remains elusive due to the strong oxidizing power of the pentazolate ion \cite{yang2022ionic}. So far, no evidence of cyclo-N$_5$$^{+}$ has been detected in any gas phase, solution, or solid form. Additionally, the N$_5$$^{+}$N$_5$$^{-}$ salts have long been desired as potential ultrahigh-performing explosives or propellants \cite{fau2002stability}. However, theoretical calculations showed that pure N$_5$$^{+}$N$_5$$^{-}$ salt is thermodynamically unstable and prone to spontaneous exothermic decomposition \cite{dixon2004enthalpies}. In this work, we address these challenges via a new strategy to attain cyclo-N$_5$$^{+}$ cation and stabilize N$_5$$^{+}$N$_5$$^{-}$ salt assembly in tailored ionic compounds. The fundamental concept behind this strategy is the utilization of negative/positive host ions as oxidizing/reducing agents to form cyclo-N$_5$$^{+}$/N$_5$$^{-}$ species. Employing advanced crystal structure search CALYPSO methods combined with \textit{ab initio} molecular dynamics (AIMD) simulations, we identified XN$_5$N$_5$F (X = Li, Na, and K) compounds that are stable at high pressures and recoverable under ambient atmosphere conditions. The present approach opens new prospects for the formation and stabilization of N$_5$$^{+}$N$_5$$^{-}$ salt assembly within ionic compounds, and this strategy may find wide adoption in creating and developing materials with tailored properties.

\section{Computational methods}
We performed structure searches for XN$_5$N$_5$F (X = Li, Na, and K) with 1–4 formula units using the advanced CALYPSO method \cite{wang2010crystal,wang2012calypso,gao2019interface,shao2022symmetry} by treating the N$_5$ ring as a molecular unit. Structure relaxations were performed using density functional theory \cite{kohn1965self} as implemented in the VASP code \cite{kresse1996efficient}. The exchange-correlation potential was described by the Perdew–Burke–Ernzerhof (PBE) generalized gradient approximation \cite{perdew1996generalized}, and electron-ion interactions by the projector-augmented-wave potentials \cite{blochl1994projector}, with 2\textit{s}$^1$, 3\textit{s}$^1$, 3\textit{s}$^2$3\textit{p}$^6$4\textit{s}$^1$, 2\textit{s}$^2$2\textit{p}$^5$, and 2\textit{s}$^2$2\textit{p}$^3$ configurations as valence electrons for Li, Na, K, F, and N atoms, respectively. A plane-wave cutoff energy of 600 eV was employed, along with a Monkhorst–Pack \textit{k} mesh \cite{monkhorst1976special} spacing of 0.25 Å$^{-1}$ for Brillouin zone sampling. These settings were chosen to ensure energy and force convergence to 10$^{-5}$ eV and 0.01 eV Å$^{-1}$, respectively. AIMD simulations for XN$_5$N$_5$F (X = Li, Na, and K) were performed using the Nosé–Hoover chain thermostat for a duration of 10 ps, with a time step of 1 fs, and at a temperature of 300 K. The 2 $\times$ 1 $\times$ 1 supercell employed in the simulation comprises a total of 96 atoms, containing eight XN$_5$N$_5$F molecular formulas.

\section{Results and discussion}

The realization of cyclo-N$_5$$^{+}$ cation requires having an element with a stronger oxidizing power than that of the pentazolate ion. Halogen elements, particularly F, are known for their electron-capturing ability, which suggests their potential capability to form the elusive cyclo-N$_5$$^{+}$ cation. To pursue this idea, we constructed a theoretical model for the isolated XN$_5$ molecule, where X denotes alkali metal elements (Li, Na, K) that have low ionization potential and halogen elements (F, Cl, Br) with strong affinity for electrons. We used a Bader charge analysis to examine electron transfer and interaction between cyclo-N$_5$ and X elements. Results (Fig. 1) show that cyclo-N$_5$ accepts electrons from alkali metals, resulting in the formation of the cyclo-N$_5$$^{-}$ anion, which is consistent with the observation of alkali metal pentazolate salts. Additionally, cyclo-N$_5$ exhibits stronger oxidizing power than Cl and Br. However, cyclo-N$_5$ exhibits electron-donating ability when bonded to F, as evidenced by the transfer of 0.23 $\left | e \right |$ from cyclo-N$_5$ to F, resulting in the formation of cyclo-N$_5$$^{+}$ cation. This result suggests that highly ionic compounds such as alkali metal fluorides could provide the necessary conditions for formation and stability of N$_5$$^{+}$N$_5$$^{-}$ salts, where cyclo-N$_5$$^{+}$/N$_5$$^{-}$ ions form via electron donating/accepting from host ions.

\begin{figure}[htp]
\centering
  \includegraphics[width=0.85\linewidth,angle=0]{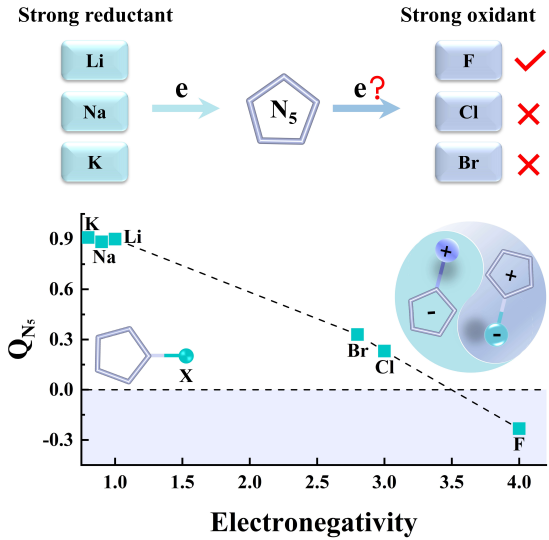}
  \caption{\label{fig:1} Schematic illustration of the design strategy for assembling cyclo-N$_5$$^{+}$ and cyclo-N$_5$$^{-}$ in tailored ionic compounds. \textit{Q}$_{N_5}$ represents accepting electrons by cyclo-N$_5$ from X elements based on the hypothetical model of isolated XN$_5$ molecules, where X refers to Li, Na, K, Br, Cl, and F atoms.}
\end{figure}

\begin{figure*}
    \centering
    \includegraphics[width=0.9\linewidth]{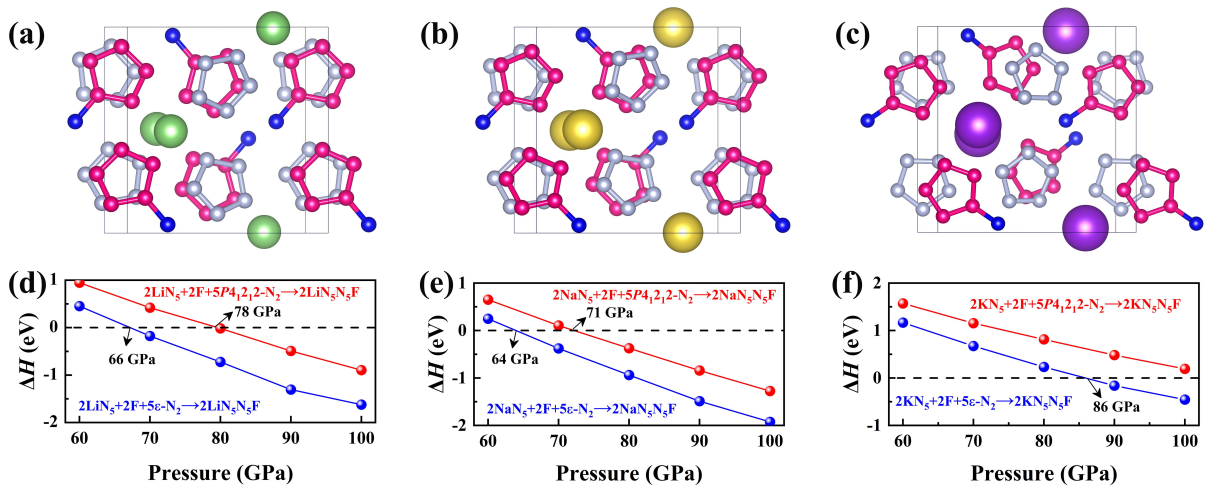}
    \caption{Crystal structures of (a) LiN$_5$N$_5$F, (b) NaN$_5$N$_5$F, and (c) KN$_5$N$_5$F at 100 GPa. Li, Na, K, N, and F atoms are represented by green, yellow, purple, grey/pink, and blue spheres, respectively. (d)-(f) Calculated enthalpy as a function of pressure of XN$_5$N$_5$F (X = Li, Na, and K) compounds relative to the experimentally found alkali metal pentazolate salts. Negative relative enthalpy indicates the stability of the XN$_5$N$_5$F compounds.}
    \label{fig:2}
\end{figure*}

To identify the stable phase assembled by cyclo-N$_5$$^{+}$/N$_5$$^{-}$ ions, we improved the CALYPSO method by treating N$_5$ ring as a molecular unit. The structure search identified three stable XN$_5$N$_5$F (X = Li, Na, and K) compounds with space group \textit{P}2$_1$2$_1$2$_1$ at 100 GPa. As shown in Figs. 2(a)-(c), these phases exhibit similar structural characteristics and bonding patterns, which contain two types of N$_5$ rings highlighted by different (pink and grey) colored atoms. The average lengths of the N-N bonds within both types of N$_5$ rings are nearly equal, at $\sim$1.28 Å. The F atoms preferentially bond to one type of N$_5$ rings. The N-F bond lengths in the three compounds show slight variations: $\sim$1.30 Å in LiN$_5$N$_5$F, $\sim$1.31 Å in NaN$_5$N$_5$F, and $\sim$1.29 Å in KN$_5$N$_5$F.

\begin{figure}
    \centering
    \includegraphics[width=0.9\linewidth]{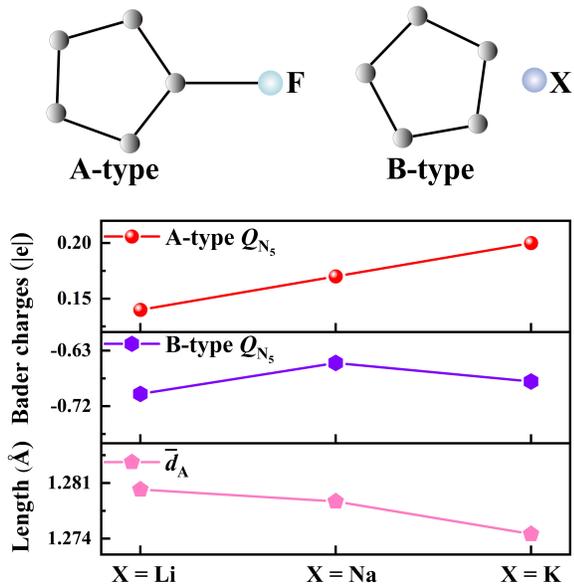}
    \caption{(a) N$_5$F and XN$_5$ units in XN$_5$N$_5$F compounds. Cyclo-N$_5$ species in the N$_5$F and XN$_5$ units are designated as A-type and B-type, respectively. (b) Bader charges of A-type and B-type cyclo-N$_5$ in XN$_5$N$_5$F (X = Li, Na, and K) compounds at 100 GPa.}
    \label{fig:3}
\end{figure}

\begin{figure*}
    \centering
    \includegraphics[width=0.9\linewidth]{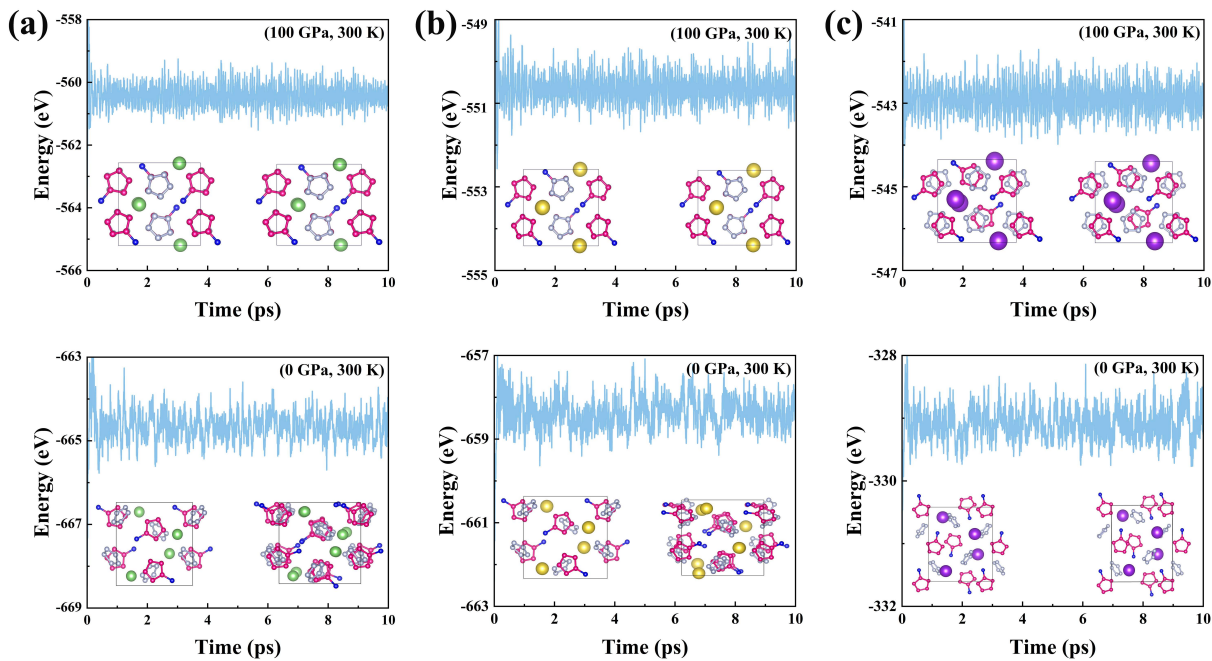}
    \caption{Evolution of energies for (a) LiN$_5$N$_5$F, (b) NaN$_5$N$_5$F, and (c) KN$_5$N$_5$F at 100 GPa, 300 K and 0 GPa, 300 K during the 10-ps AIMD simulations. The insets feature structural snapshots captured at the beginning and end of the 10-ps simulation period.}
    \label{fig:4}
\end{figure*} 

\begin{figure}
    \centering
    \includegraphics[width=0.9\linewidth]{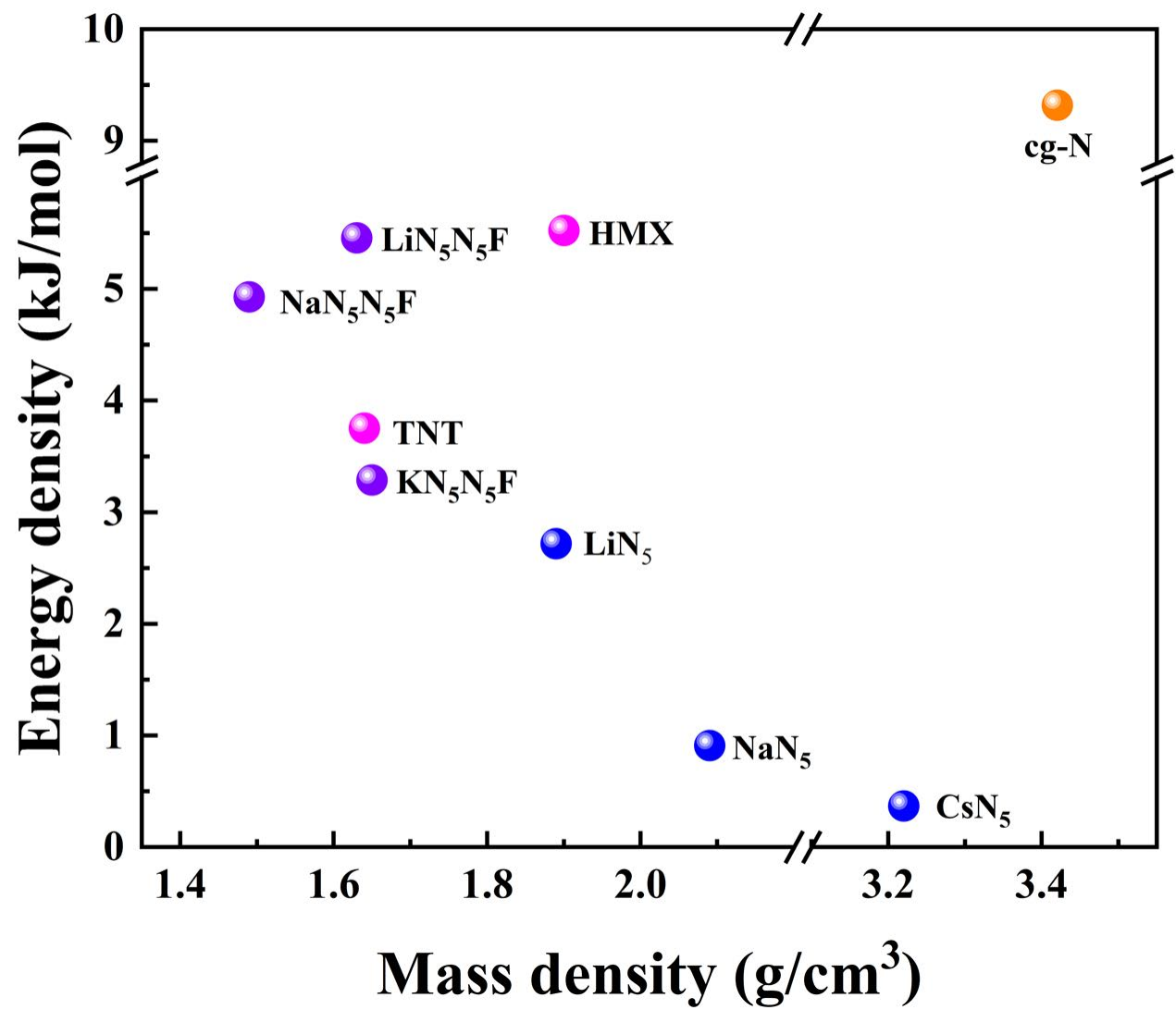}
    \caption{Energy density \textit{versus} mass density for XN$_5$N$_5$F, compared to various alkali-metal pentazolates, cg-N, and the well-established high-energy density materials HMX and TNT.}
    \label{fig:5}
\end{figure}

To evaluate the stability of XN$_5$N$_5$F compounds, we calculated their formation enthalpies with respect to decomposition at different pressures, adopting \textit{P}2$_1$/\textit{c}-LiN$_5$ (60 GPa) and \textit{C}2\textit{c}-LiN$_5$ ($>$60 GPa) \cite{shen2015novel}, \textit{Cm}-NaN$_5$ \cite{steele2016sodium}, \textit{P}2$_1$\textit{m}-KN$_5$ \cite{steele2017novel}, $\varepsilon$-N \cite{olijnyk1990high}, \textit{P}4$_1$2$_1$2-N \cite{pickard2009high}, and \textit{Cmca}-F \cite{lv2017crystal} as reference structures. Here $\varepsilon$-N is experimentally found molecular solid nitrogen and \textit{P}4$_1$2$_1$2-N is theoretically determined lowest-enthalpy molecular solid nitrogen. Our calculations show [Figs. 2(d)-(f)] that when considering decomposition into experimentally found $\varepsilon$-N, LiN$_5$N$_5$F and NaN$_5$N$_5$F become energetically stable at relatively lower pressures of 66 and 64 GPa, respectively, while KN$_5$N$_5$F reaches stability at a higher pressure of 86 GPa. However, when considering decomposition into the theoretical lowest-enthalpy molecular \textit{P}4$_1$2$_1$2-N, the pressures for LiN$_5$N$_5$F and NaN$_5$N$_5$F to have negative formation enthalpies increase into 78 and 71 GPa, respectively, while KN$_5$N$_5$F becomes unstable against the decomposition. As the radius of the X element in XN$_5$N$_5$F compounds increases, the pressure required to stabilize XN$_5$N$_5$F gradually increases. Therefore, larger-sized alkali metal elements such as Rb and Cs are not considered in the calculations due to the expected higher pressures for the stabilization, if achievable at all, of their corresponding compounds. 

Bader charge analysis is widely recognized and utilized for partitioning electron density distribution into atomic regions. It is regarded as a physically unbiased and robust method for determining charges in compounds, offering valuable insights into the distribution of electronic charges within a given system \cite{oganov2009ionic}. In our analysis, we estimate the charge transfer in XN$_5$N$_5$F compounds based on a systematic Bader charge analysis. As depicted in Fig. 3(a), the two unequal types of cyclo-N$_5$ rings in XN$_5$N$_5$F compounds are designated as A-type and B-type cyclo-N$_5$, which are associated with the N$_5$F and XN$_5$ structural units, respectively. The Bader charge analysis [Fig. 3(b)] reveals that electron transfer takes place from A-type cyclo-N$_5$ to F atom due to the stronger oxidizing power of F compared to cyclo-N$_5$. Specifically, A-type cyclo-N$_5$ species in LiN$_5$N$_5$F, NaN$_5$N$_5$F, and KN$_5$N$_5$F loses 0.14, 0.18, and 0.20 $\left | e \right |$, respectively, resulting in cyclo-N$_5$$^{+}$ cations in these compounds. Conversely, B-type cyclo-N$_5$ species within XN$_5$N$_5$F compounds experience electron transfer from the X (Li, Na, or K) atoms to the cyclo-N$_5$, leading to the formation of cyclo-N$_5$$^{-}$ anions. Therefore, the cyclo-N$_5$$^{+}$ cations, formed by electron transfer to F atoms, and the cyclo-N$_5$$^{-}$ anions, formed by receiving electrons from X atoms, can be assembled in the environment of highly ionic compounds XF.

We performed AIMD simulations to assess thermodynamic stability of XN$_5$N$_5$F compounds. The results, as depicted in Figs. 4(a)-(c), indicate that the energy oscillation of XN$_5$N$_5$F stays in a small energy range of $\sim$1 eV per 2 $\times$ 1 $\times$ 1 supercell, equivalent to $\sim$ 10 meV/atom at 100 GPa and 300 K. A comparison between the initial and final structures reveals well-preserved molecular unit and overall crystal structure, suggesting that XN$_5$N$_5$F compounds well maintain their integrity without significant structural changes under the examined temperature-pressure conditions.  

To check for possible metastability of XN$_5$N$_5$F at ambient conditions, we further carried out AIMD simulations at 0 GPa and 300 K. The resulting energy oscillation of XN$_5$N$_5$F increases slightly, measuring $\sim$20 meV/atom, but remains in an acceptable range. In particular, the cyclo-N$_5$ ions in XN$_5$N$_5$F only undergo mild spinning and twisting motions. These dynamic movements do not compromise the structural integrity of the cyclo-N$_5$ ions, as they remain intact without decomposition and largely stay in their original crystal positions. These results show that once XN$_5$N$_5$F compounds are synthesized at high pressures, they hold the potential to be recoverable to ambient atmospheric conditions.

Given the feasibility of XN$_5$N$_5$F compounds as metastable phases at ambient atmospheric conditions, we further assessed their potential as high-energy density materials by calculating the decomposition energy. Total energy calculations were performed for XN$_5$N$_5$F compounds relative to XF ionic compounds and $\alpha$-N$_2$. The results indicate that LiN$_5$N$_5$F, NaN$_5$N$_5$F, and KN$_5$N$_5$F can release energies in the amounts of 0.79, 0.78, and 0.57 eV/atom, respectively. These values correspond to energy densities of 5.46, 4.93, and 3.29 kJ/g. In Fig. 5, a comparison of XN$_5$N$_5$F compounds with alkali pentazolate salts and well-established high-energy density materials reveals not only significantly larger energy densities but also lower mass densities. Specifically, the energy density of LiN$_5$N$_5$F is 5.46 kJ/g,  which is comparable to that of HMX (5.53 kJ/g) and larger than that  of TNT (3.76 kJ/g) \cite{zhang2016high}. Notably, the mass density of LiN$_5$N$_5$F is 1.63 g/cm$^{3}$, which is smaller than that (1.90 g/cm$^{3}$) of  HMX. Although the energy densities of XN$_5$N$_5$F compounds are lower than that of cg-N (9.70 kJ/g) \cite{li2018route}, they exhibit signifi-cantly lower mass densities ranging from 1.49 to 1.65 g/cm$^{3}$, which are approximately half of the mass density of cg-N (3.42 g/cm$^{3}$). This suggests that XN$_5$N$_5$F compounds offer a favorable combination of high energy release and reduced mass density achievable under ambient conditions. Such characteristics are highly desirable for the development of high-performance energetic materials.

\section{Conclusion}
In summary, we have devised a new strategy to design cyclo-N$_5$$^+$ cation and assemble long-sought N$_5$$^{+}$N$_5$$^{-}$ salts in highly ionic compounds following an electronegativity co-matching rule. Utilizing CALYPSO crystal prediction method and AIMD simulations, we identify XN$_5$N$_5$F compounds that are viable for high-pressure synthesis and remain metastable at ambient atmospheric conditions. XN$_5$N$_5$F compounds exhibit desirable characteristics of high energy release and low mass density, making them superior high-performance energetic materials. These findings open exciting prospects of realizing the N$_5$$^{+}$N$_5$$^{-}$ salt assembly in ionic compound environments, which enriches novel polynitrogens, with broad implications for the synthesis of diverse species via the targeted design of composition and charge state. 

\section{Acknowledgments}

The authors acknowledge funding support from National Key Research and Development Program of China (Grant No. 2022YFA1402304), National Natural Science Foundation of China (Grants No. T2225013, No. 12374010, No. 12074154, No. 12174142, and No. 11904142), and Open Project of State Key Laboratory of Superhard Materials, Jilin University (Grant No. 202417). The computational resources have been provided by the High-Performance Computing Center of School of Physics and Electronic Engineering of Jiangsu Normal University and High-Performance Computing Center of Jilin University.

\bibliography{references}

\end{document}